\documentclass{article}
\usepackage{graphicx}

\usepackage{graphics}
\usepackage{amsmath}
\usepackage{amssymb}
\usepackage{mathrsfs} 
\usepackage{subfigure} 
\usepackage[squaren]{SIunits} 

\let\realverbatim=\verbatim
\let\realendverbatim=\endverbatim
\renewcommand\verbatim{\par\addvspace{6pt plus 2pt minus 1pt}\realverbatim}
\renewcommand\endverbatim{\realendverbatim\addvspace{6pt plus 2pt minus 1pt}}
\makeatletter
\newcommand\verbsize{\@setfontsize\verbsize{10}\@xiipt}
\renewcommand\verbatim@font{\verbsize\normalfont\ttfamily}
\makeatother

\newcommand\Imag{\mbox{Im}} 

\newsavebox{\astrutbox}
\sbox{\astrutbox}{\rule[-5pt]{0pt}{20pt}}

\title{Instability and morphology of polymer solutions coating a fiber\footnote{Accepted in Journal of Fluid Mechanics.}}

\author{F. Boulogne, L. Pauchard, F. Giorgiutti-Dauphin\'e\footnote{Univ Pierre et Marie Curie-Paris 6, Univ Paris-Sud, CNRS, F-91405, Lab FAST, Bat 502, Campus Univ, Orsay, F-91405, France.}}

\begin{document}

\label{firstpage}
\maketitle

\begin{abstract}

We report an experimental study on the  dynamics of a thin film of polymer solution coating a vertical fiber. The liquid film has first a constant thickness and then undergoes the Rayleigh-Plateau instability which leads to the formation of sequences of drops, separated by a thin film, moving down at a constant velocity. Different polymer solutions are used, i.e. xanthan solutions and polyacrylamide (PAAm) solutions. These solutions both exhibit shear-rate dependence of the viscosity, but for PAAm solutions, there are strong normal stresses in addition of the shear-thinning effect. 
We characterize experimentally and separately the effects of these two non-Newtonian properties on the flow on the fiber. Thus, in the flat film observed before the emergence of the drops, only shear-thinning effect plays a role and tends to thin the film compared to the Newtonian case. The effect of the non-Newtonian rheology on the Rayleigh-Plateau instability is then investigated through the measurements of the growth rate and the wavelength of the instability.
Results are in good agreement with linear stability analysis for a shear-thinning fluid.  The effect of normal stress can be taken into account by considering an effective surface tension which tends to decrease the growth rate of the instability. Finally, the dependence of the morphology of the drops with the normal stress is investigated and a simplified model including the normal stress within the lubrication approximation provides good quantitative results on the shape of the drops. 
\end{abstract}

 
\tableofcontents

\section{Introduction}

The  fiber coating process  is widespread in numerous industrial applications such as the manufacture of glass, polymeric and optical  fibers, conducting cables or textile fibers. The application of a thin layer on these solid substrates should ensure mechanical or optical properties of the final deposit. Hence, it is of crucial interest to control the final thickness of the liquid film.  It has been well known since \cite{Rayleigh1878} that a cylindrical free surface of a fluid is unstable under the action of the surface tension. Later, \cite {Boys1959} described in his monograph  the patterns observed in a spider web with a sticky fluid. He reproduced the experiment with castor oil and a quartz fiber, and detailed the spatial variations of the film thickness.
For a fiber drawn out of a bath, \cite{Goucher1922} and later \cite{White1966} provided a first view of the flowing regime as a function of the capillary number and the Goucher number defined as the 
ratio between the radius of the fiber and the capillary length.  
Subsequently, different configurations have been studied to identify the different mechanisms responsible for destabilization of the film and the dependence on various parameters such as the radius of the fiber, the viscosity, the inertial forces or the influence of surfactant on the growth rate of the instability  \cite[]{Goren1962, Goren1964, Quere1990,Quere1999,Carroll1974, Duprat2009c, Duprat2009b, Smolka2008, Kliakhandler2001}. More recently, we have reported a flow regime diagram which identifies, depending on the fiber radius and the flow rate on the fiber, the dominant physical mechanisms \cite[]{Duprat2009b, Duprat2007,Duprat2009c}. 
It appears that for small fiber radii compared to the capillary length and low flow rates, the liquid film is dominated by the surface tension, and the instability mechanism is the Rayleigh-Plateau instability whose nature is absolute. 
For higher values of the parameters (fiber radius and/or flow rate) two other regimes have been discerned respectively dominated by gravity and drag (the drag/dravity regime) or by inertia and drag (the drag/inertia regime). The nature of the instability is then convective for these regimes. 
Extensive theoretical studies have investigated  the dynamics of the film
\cite[]{Frenkel1987, Quere1990, Kalliadasis1994}. Among the most recent studies, we can mention the works of \cite{Kliakhandler2001} and \cite{Craster2009} on film thickness of the same order as the radius fiber and with negligible inertia contribution. Their numerical results are compared with experimental results \cite[]{Duprat2009b} performed in the three kinds of regime depicted above with a predictable deviation for the Drag/Inertia regime.

In most industrial situations, the coating fluid is a polymer material or a complex fluid and exhibits non-Newtonian properties depending on miscellaneous parameters such as concentration, structure or flexibility of polymers.  Yield stress, shear-thinning or elastic effects  are some of the non-Newtonian behaviors which can affect the structure of the flow,  the appearance of the instability or the morphology of the patterns. \cite{DeRyck1998}, in the case of a ``dip-coating''  configuration where the fiber is drawn out of a bath of liquid, observed that the film swells due to the presence of polymer in the solution. Considering the normal stress and the lubrication approximation, they found an analytical expression of the film thickness as a function of the withdrawal velocity and the normal stress coefficient. In general for most of the cases, when the instability is studied for non-Newtonian fluids it reveals different classes of patterns and the presence of polymers can drastically 
change the dynamic of the system, as in Faraday or Saffman-Taylor instabilities \cite[]{Lindner2009}. In the case of the instability of a liquid jet \cite[]{Clasen2006,Wagner2005,Eggers2008,Bhat2010}, the addition of polymers causes the formation of a ``beads-on-a-string'' structure where adjacent beads are joined by a thread which grows thinner and  strongly delays the detachment of droplets. 
In this configuration, the flow is subject to a strong elongation: a velocity gradient exists in the direction of the flow due to gravity forces. This additional resistance to breakup compared to a simple fluid is due to large extensional stresses. Contrary to the case of a liquid jet, there is no elongational viscosity in the case of the flow down a fiber due to the no-slip condition on the fiber as detailed further. 

A complication inherent in the use of these complex fluids is that they exhibit different non-Newtonian properties with opposite effects. Notably, most polymer solutions are both shear-thinning  and present elastic effects. 
In the context of fiber coating, and to obtain independently the role of the shear-thinning effect and normal stress on the Rayleigh-Plateau instability,  we have performed experiments with two different polymer solutions: one with a rod-like polymer (xanthan), exhibiting a pure shear-thinning effect; the other with a flexible polymer (polyacrylamide abbreviated as PAAm) which exhibits non-negligible normal stress along with shear-rate dependence of viscosity similar to xanthan solutions.

Our experiments are all performed in a regime where inertial and gravitational forces are negligible compared with capillary, elastic and viscous forces. The Rayleigh-Plateau instability is then absolute~\cite[]{Duprat2007} and the flow patterns consist of drops, where fluid is partly trapped in a recirculation zone, sliding down a very thin (smaller than $100$ $\mu$m) and quasistatic liquid substrate. This droplike wave train emerges from a constant film thickness, i.e. the flat film region. We investigate the role of the non-Newtonian properties on such a flow, i.e, the flat film and the drop-like wave train resulting from the Rayleigh-Plateau instability.
\\
This paper is organized as follows. 
In section \ref{sec:sample}, we proceed to a rheological characterization of the solutions. In section~\ref{sec:matos}, we present the experimental setup and visualisation techniques. In section~\ref{sec:flat}, we look at the flat film before the appearance of the instability and the shear-thinning effect on the film thickness. In section~\ref{sec:insta}, the experimental growth rate and the wavelength of the instability are measured experimentally for xanthan solutions (only shear-thinning effect) and for PAAm solutions exhibited strong normal forces and similar viscosity shear-rate dependence. These experimental data are then compared to the results of a linear stability analysis taking into account a non constant viscosity with the shear-rate. In the last part, section~\ref{sec:drop}, we will treat the effect of normal stress on the morphology of the drops and provide a simplified model to explain the dependence of the drop shape with the normal stress.

\section{Materials and characterizations}
\label{sec:matos}
Before considering the non-Newtonian properties, certain conditions are required: first, to avoid inertial forces, and second to ensure a perfect wetting on the fiber. In a previous paper \cite[]{Duprat2009b}, we have 
presented a diagram of the expected flow regime which details the dominant physical mechanisms in the plane of the dimensionless numbers $R/l_{c}$ versus $h_{0}/R$ with $R$, the radius of the fiber, $l_{c}$ the capillary number and $h_{0}$ the flat film thickness. To ensure that the flow is dominated by capillary forces, with no inertia, some conditions on the fiber radius, the surface tension and the viscosity of the fluid must be fulfilled in agreement with the flow regime diagram mentioned above. In that capillary region, the flow on the fiber consists of drops sliding on a quasistatic thin film. To ensure a good wetting on the fiber and consequently an axisymmetric pattern on the fiber, the surface tension has to be lower than $40$ mN/m. Finally, these two conditions are satisfied by using a fiber radius equal to $0.28$ mm and solutions composed of a mixture of water, glycerol (to increase viscosity) and surfactant to reach a surface tension close to $30$ mN/m.

Some previous experiments carried out with Newtonian fluids on the same experiment \cite[]{Duprat2009b}, have shown that film thicknesses between $0.1$ and $1$ mm are possible with velocities ranging from $1$ to $10$ cm/s and viscosities between $50$ and $500$~mPa.s. This implies that the shear-rate range encountered in our experiments is from $10$ to $1000$ s$^{-1}$.

The liquids used are then semi-dilute solutions of polymers: xanthan and  Polyacrylamide (PAAm) purchased from Sigma-Aldrich. Both present shear-thinning behavior but only the second one exhibits large elastic effects in the shear-rate range considered in the experiments. To discern the effect of elasticity, Boger fluids would have been theoretically more appropriate. Nevertheless, in practice, the large quantities of liquid required for our experiments and the possible degradation of the nylon fiber by the solvents used in Boger fluids, are two reasons prohibiting their use.

\subsection{Samples preparation}
\label{sec:sample}
\subsubsection{Polyacrylamide}

PAAm is a high molecular weight chain resulting in long flexible chains ($M \simeq 1.5 \times 10^7$ g/mol).
The samples are prepared very carefully to get homogeneous solutions. 
First, a solvent is prepared by mixing $50$\% of purified water and $50$\% of pure glycerol (all percentages presented in this article are weight percentages).
Six polymer concentrations were studied in the range from $0.1\%$ to $0.6\%$.
The surfactant selected to reduce the surface tension is Triton X-100 (TX-100). This choice was motivated by its mixing properties at high polymer concentrations and for the resulting low surface tension~\cite[]{Zhang1998}.
The TX-100 concentration is $4.5$\% (about $300$ times the Critical Micelle Concentration, CMC, in water).
For this high concentration we assume that the surfactant mobility timescale is higher than the timescale for the instability growth rate. Indeed, the time variation of the surface tension due to diffusion of surfactant is $10^{-2}$ s for a TX-100 concentration $50$ times the CMC in water \cite{Fainerman1994}. 
This time scale is of the same order of magnitude as the characteristic growth rate of the instability:
$\frac{\eta R^4}{\gamma h^3} \sim \frac{0.1\times(0.6\times10^{-3})^4}{30\times10^{-3}(0.2\times10^{-3})^3} = 0.05$ s (using typical values for the viscosity $\eta$, the fiber radius $R$, the surface tension $\gamma$, and the film thickness $h$).
Consequently, in the range of concentrations considered in our experiments we assume that the interface is rapidly saturated with surfactant molecules before the instability occurs.
The surface tension of the final solution is then $\gamma = 32.3\pm0.5$ mN/m. 
We should note that no apparent rheological modifications are observed by varying the TX-100 concentration. 

\subsubsection{xanthan}

xanthan is a rigid rod-like polymer ($M \simeq 5 \times 10^6$ g/mol). 
A preparation protocol similar to the PAAm solutions was used for xanthan solutions. 
The solvent is slightly different: $60$\% of glycerol and $40$\% of water. 
Since xanthan is a polyelectrolyte polymer, the resulting rheological properties of this polymer are known to be modified by the addition of salt \cite[]{Wyatt2009}. 
Thus, different concentrations of NaCl  allow adjustment of the rheological properties.
After the addition of TX-100,  the surface tension of the final solution is $\gamma=32.7\pm0.8$ mN/m independent of the salt concentration.

\subsection{Rheological characterization}
The rheology of polymer solutions was performed using an ``Anton Paar'' rheometer with a cone and plate geometry.
We have chosen a large cone (radius: $49.988$ mm) with small angle (angle: $0.484^\circ$) in order to measure precisely normal stress in a large range of shear-rates.
The sample temperature was fixed at $20.00\pm0.05^\circ$C. 

\subsubsection{Polyacrylamide}
\begin{figure}
	  \begin{center}
	  \includegraphics[scale=1]{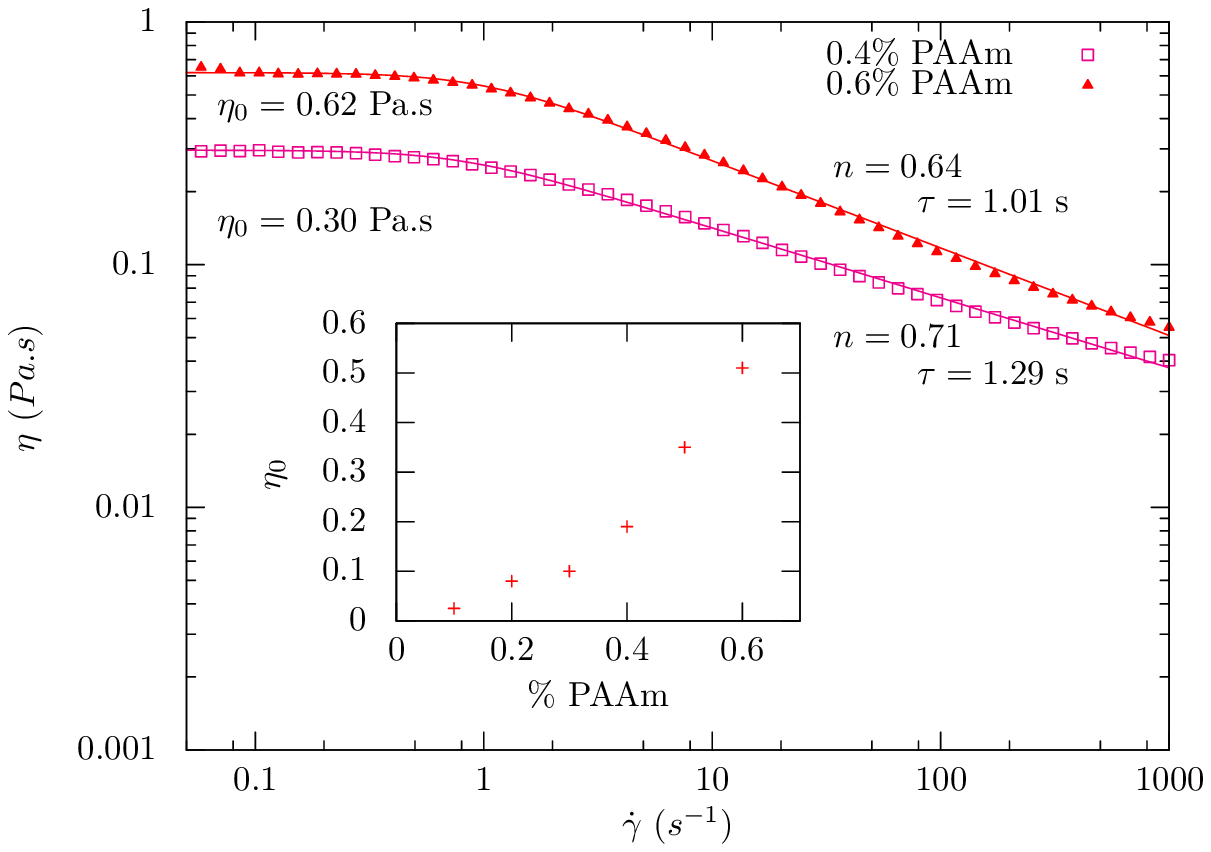}
	  \caption{Variations of the shear viscosity $\eta$, of PAAm solutions vs. shear-rate $\dot\gamma$, in a log-log scale. 
	  Each symbol refers to a different polymer concentration. 
	  Plain curves correspond to data fits with a Carreau model. The inset shows the zero shear-rate viscosity $\eta_0$, as a function of the concentration in PAAm.}
	  \label{fig:visco}
	  \end{center}
\end{figure}

The evolution of the apparent shear viscosity, $\eta$, is plotted versus the shear-rate, $\dot{\gamma}$, for two concentrations in figure \ref{fig:visco}.
As usual, $\eta$ decreases with $\dot{\gamma}$ and, at a given shear-rate, increases with the polymer concentration.
Such curves are typical for shear-thinning fluids where a constant Newtonian low-shear viscosity is followed by a power-law dependence before reaching the viscosity of the solvent at high shear-rates.
These measurements can be reasonably fitted by a four-parameter Carreau model~\cite[]{Macosko1994}:

\begin{equation}
	\eta = \eta_{\infty} + \frac{\eta_{0}-\eta_{\infty}}{ \left(1+ (\tau \dot\gamma)^2 \right)^{\frac{1-n}{2}}  }
	\label{eq:carreau_law}
\end{equation}

where $\eta_{0}$ and $\eta_{\infty}$ are respectively the viscosity for the zero-shear limit and infinite shear-rate, and $\tau$ denotes a characteristic time scale that measures the scale at which the shear-thinning effect becomes important. 
The exponent $n$ is the power of the following Ostwald power-law equation:

\begin{equation}
	\eta= \beta\dot\gamma^{n-1}\label{eq:power_law}
\end{equation}

The zero-shear limit increases rapidly with the polymer concentration, that is typical of entangled polymer solutions (see the inset of figure \ref{fig:visco}).
Recently, it has been shown~\cite[]{Liu2009} that in a good solvent the entangled concentration $c_e$ for PAAm is about nine times the crossover concentration $c^*\simeq 0.2$ g/L.
The temperature effect on the samples indicates that the viscosity decreases by $10$\% for a $5^\circ$C increase. \\

\begin{figure}
	  \begin{center}
	   \subfigure[]{\label{fig:n1_a}
	  \includegraphics[scale=1]{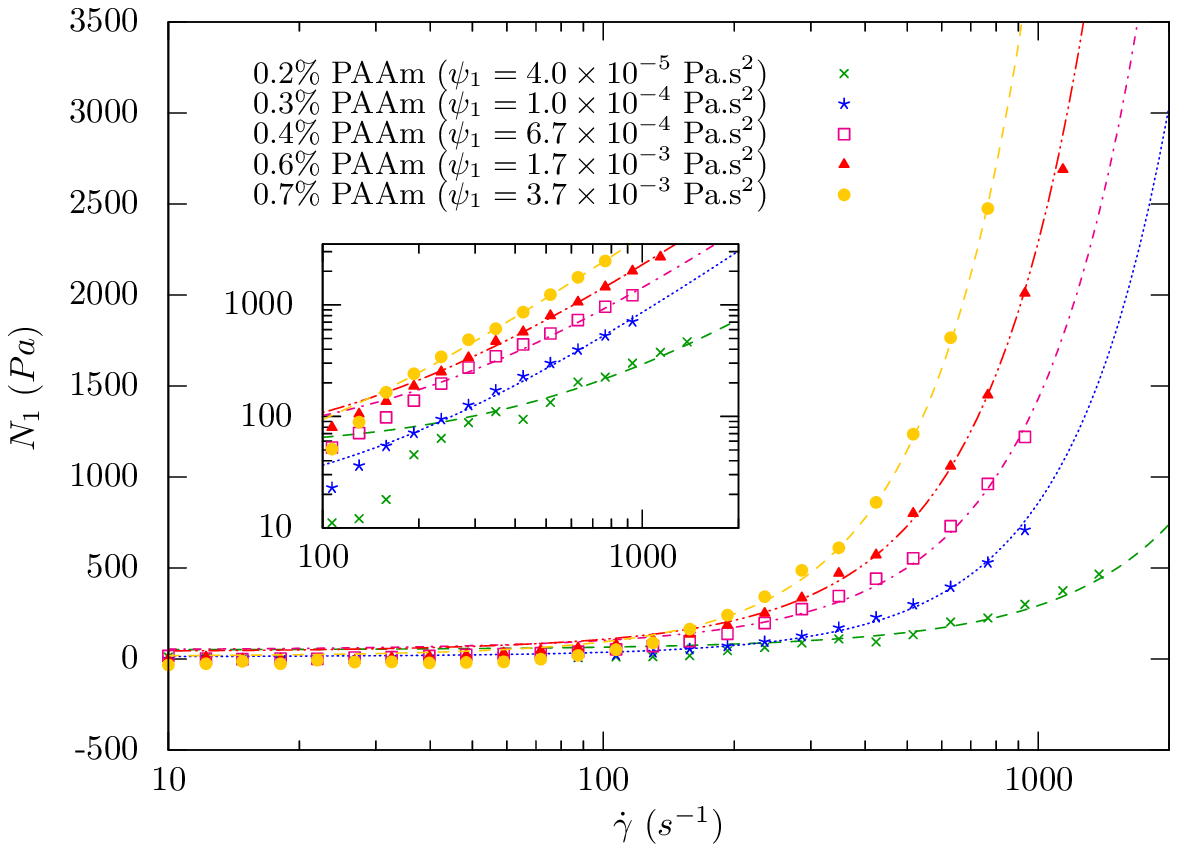}}
	 \subfigure[]{\label{fig:n1_b}
	     \includegraphics[scale=1]{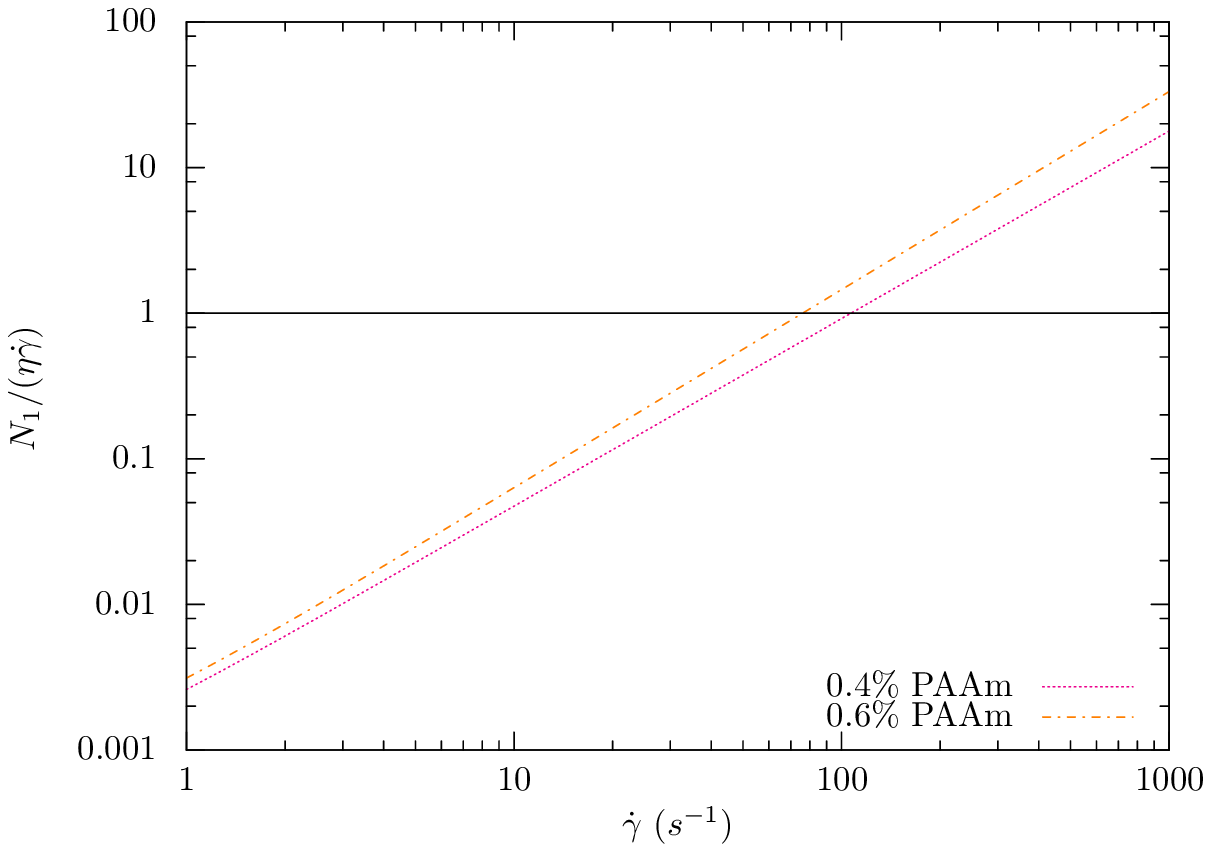}}
	      \caption{(a) Log-log plot of the normal stress, $N_1$, as a function of the shear-rate, $\dot\gamma$, for PAAm solutions in water-glycerol ($50$\% : $50$\%) solvent; $4.5$\% TX-100 surfactant was added to the solutions. Plain curves correspond to data fits using equation \ref{eq:n1_law}. (b) $\frac{N_{1}}{\eta(\dot\gamma)\dot\gamma}$ versus shear-rate, $\dot\gamma$ , for PAAm solutions}
	  \label{fig:n1}
	  \end{center}
\end{figure}

The normal stress measurements are presented in figure \ref{fig:n1_a} as a function of the shear-rate for different PAAm concentrations over a wide range of shear-rates.
A significant increase of the normal stress with the shear-rate is observed in accordance with:

\begin{equation}
	  N_1=\psi_1 \dot\gamma^2
	  \label{eq:n1_law}
\end{equation}

where $\psi_1$ is the first normal stress coefficient characterizing the fluid \cite[]{Bird1987}.
Values are given in figure \ref{fig:n1_a} for several Polyacrylamide concentrations. 
In the inset graph, we have presented the data using a double log-plot. It appears more clearly that, except for PAAm solutions at $0.6\%$, there is a discrepancy between the data and the curve fit for low shear-rates (lower than $200$ s$^{-1}$), indicating dependence of the first normal stress difference with the shear-rate. Thus the $0.6\%$ solution would be the best candidate for studying the normal force effect in section \ref{sec:drop}.
Normal stress magnitude can be compared to viscous stress by estimating the ratio $\frac{N_{1}}{\eta(\dot\gamma)\dot\gamma}$ as a function of the shear-rate \ref{fig:n1_b}. 
This ratio increases with the shear-rate, highlighting the importance of the normal stress, which starts to be dominant compared to the shear-thinning effect for shear-rates larger than $100$ s$^{-1}$. One should then expect a large amount of normal stress in the drops, for which the shear-rate is always larger than $100$ s$^{-1}$.

\subsubsection{xanthan}

\begin{figure}
\begin{center}
\includegraphics[scale=1]{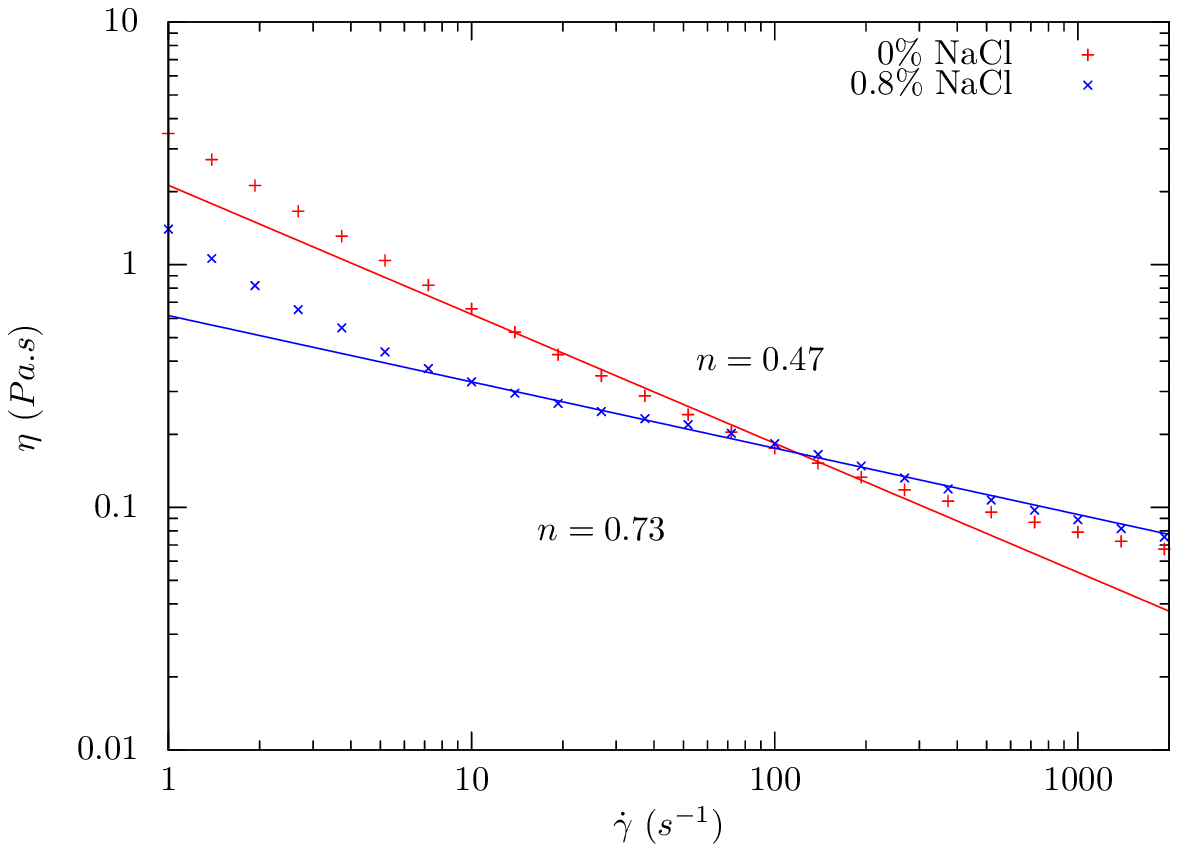}
\caption{Viscosity $\eta$,  as a function of shear-rate $\dot\gamma$, for two solutions constituted by water-glycerol~($50$\% : $50$\%), $4.5$\% TX-100 and $0.4$\% xanthan. Red cross correspond to free-salt solution and blue one are for a salt concentration up to $0.8$\%. Numbers indicate the slopes of the Ostwald power-law model from fits over a $\dot\gamma$ range: $[10;2000]$ s$^{-1}$.}
\label{fig:xanthan}
\end{center}
\end{figure}

Rheological measurements are typical of shear-thinning fluids where a constant Newtonian low-shear viscosity is followed by a power-law dependence before reaching the viscosity of the solvent at high shear-rates (figure \ref{fig:xanthan}).
Nevertheless, as shown in figure \ref{fig:xanthan}, the power-law behavior failed to fit the experimental results in the whole range of shear-rates.
Since xanthan is a polyelectrolyte, the solution rheology and molecular configuration are greatly affected by the solution's ionic strength.
Thus by adding $0.8$\% NaCl to the solution, the shear-thinning effect can be adjusted to $n=0.73$ in a reasonable range of shear-rates from $5$ to $2000$ s$^{-1}$. 
No significant normal stress has been detected for xanthan below $4000$ s$^{-1}$, a shear-rate which is not expected to be reached in the experiment. Subsequently, we will exclusively use salted xanthan solutions as pure shear-thinning solutions.

\subsection{Experimental setup}

\begin{figure}
\begin{center}
\subfigure[]{\label{fig:manip_a}
\includegraphics[scale=0.8]{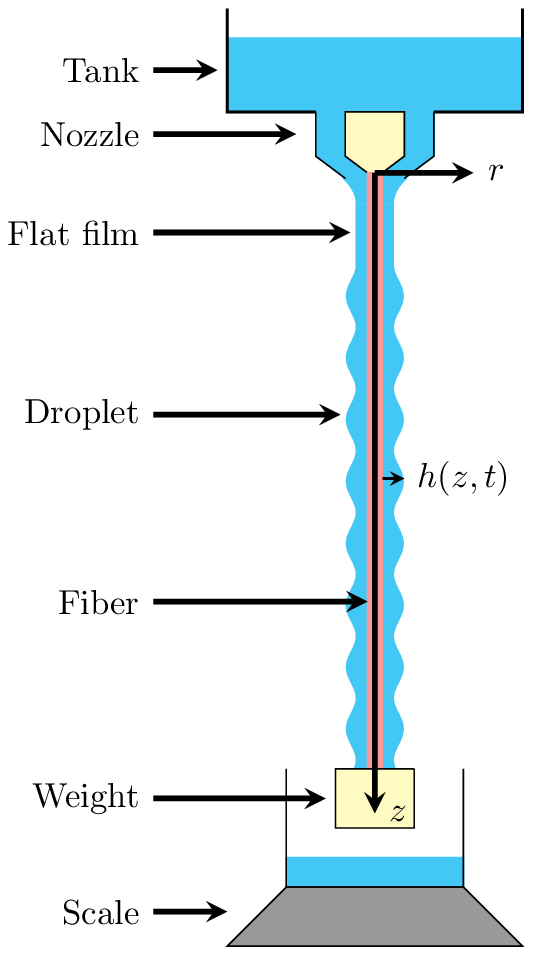}}
\subfigure[]{\label{fig:manip_c}
\includegraphics[scale=0.20]{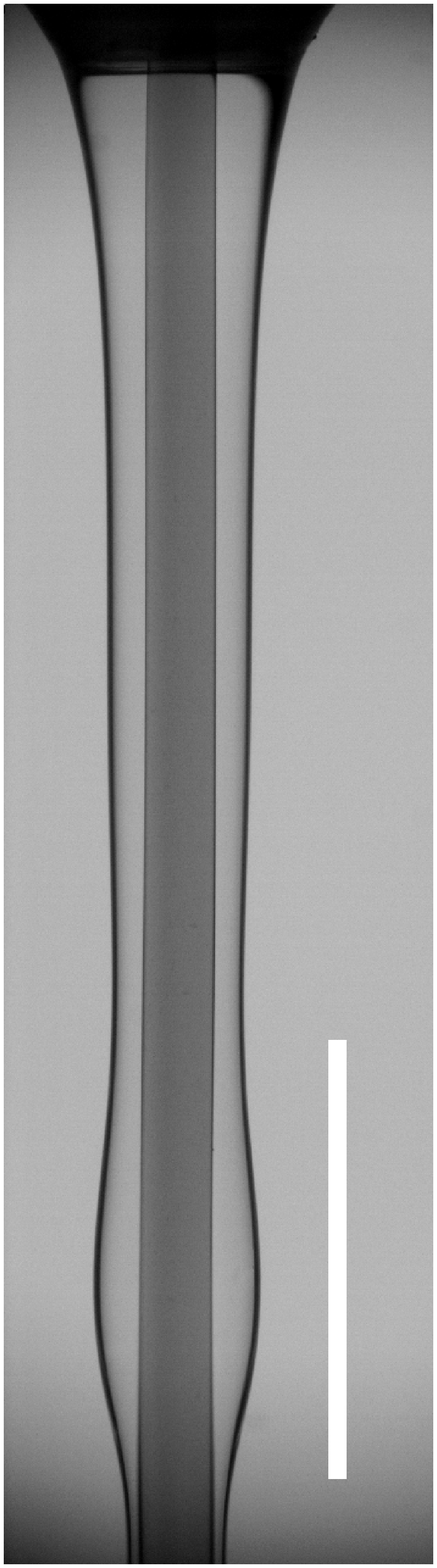}}
\subfigure[]{\label{fig:manip_b}
\includegraphics[scale=0.26]{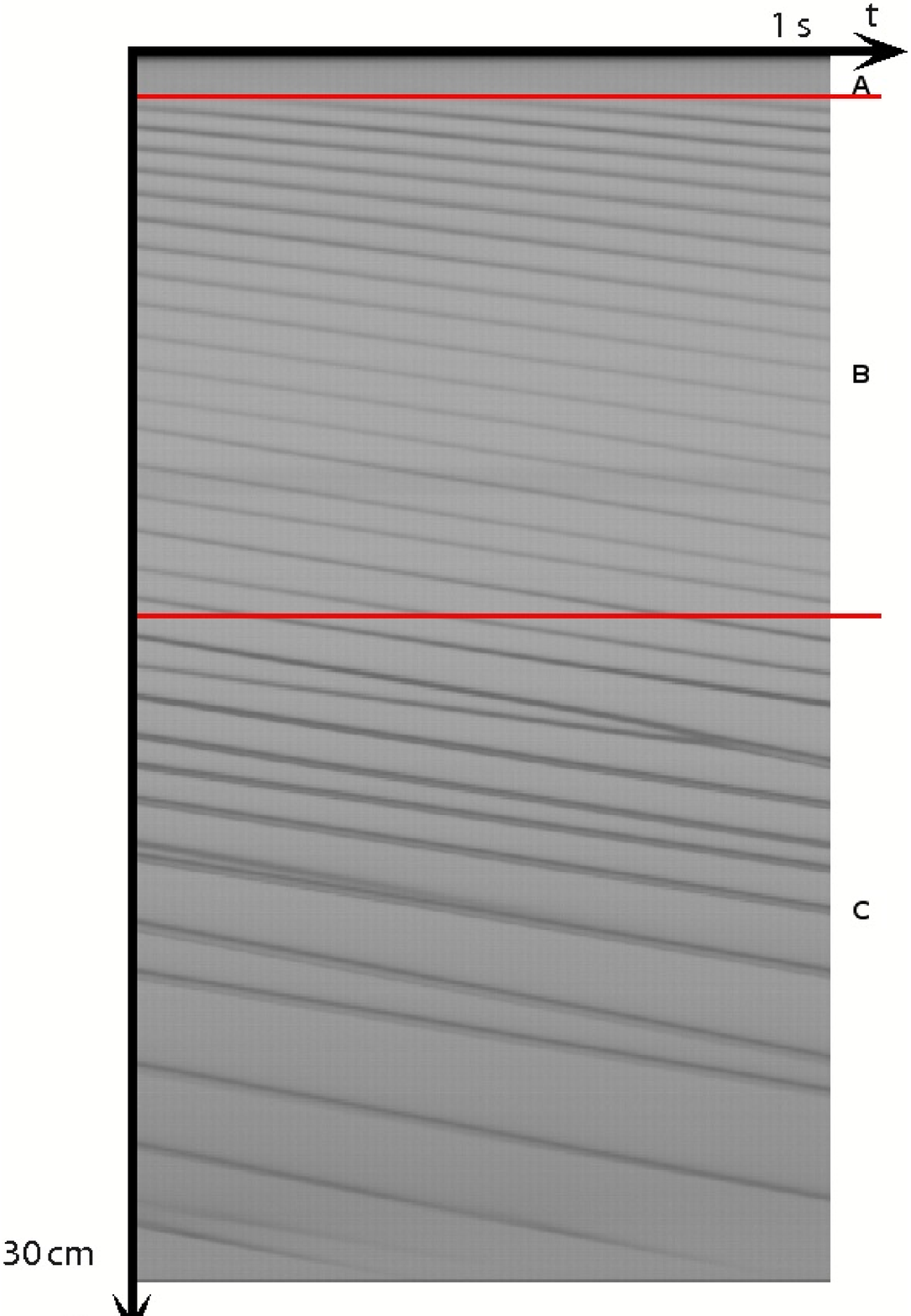}}
\caption{(a) Experimental setup. Scheme showing a fluid flowing down on a fiber from a upper tank. The flow rate is controlled by a valve and guided with a nozzle. (b) Picture snapped by a high-speed camera with a telecentric lens ($1\times$). The white bar length is $5$ mm. (c) Spatiotemporal evolution of the film obtained by a vertical linear camera passing through top drops (black lines). Letters A, B and C denote respectively the flat film, the ordered and the disordered pattern regions.}
\label{fig:manip}
\end{center}
\end{figure}

As depicted in figure \ref{fig:manip_a}, the fluid flows from an upper reservoir (diameter: $14$~cm) down a nylon fiber (diameter: $0.56$~mm). The relative pressure variation is about $0.001$\% during one minute for the highest measured flow rates. The flow rate is controlled by a valve composed of two axisymmetric cones. The mass flow rate $Q$ is measured from the weight variation of a collecting tank recorded by a computer-controlled scale.
A transparent nozzle guides the fluid on the fiber. Its verticality is crucial to obtaining an axisymmetric flow and it is ensured by a mechanical device which enables very accurate fiber displacements with a sensibility of $2.4$ arc sec. Two perpendicular cameras with zoom lens help to control the axisymmetry of the film flowing down the fiber.

As depicted in figure \ref{fig:manip_c}, the flow presents three regions along the fiber. 
A meniscus is followed by a flat film with a constant thickness on a distance called the healing length, which increases slightly with the flow rate \cite[]{Duprat2007}. Then, the Rayleigh-Plateau instability leads to the formation of a regular pattern of beads flowing on a very thin and flat film.
In this paper, we will deal only with regime dominated by capillary forces, so we exclude low flow rates (the dripping regime) and high flow rates where inertial flow dominates. The flow regime is then absolute.
The film thickness and the shape of the drops are captured by a high-speed digital camera with a telecentric lens.
The interface position is detected in both space and time, so we are able to measure the film thickness $h(z,t)$ with an accuracy of $0.02$~mm. 
A linear camera provides spatiotemporal diagrams which deliver information on the dynamic of the flow. A vertical pixel line passing through the peaks of the drops is recorded and stored at constant time intervals. The resulting spatiotemporal diagram produces the $(z,t)$ trajectories of the drops along the fiber. A typical spatiotemporal diagram is shown on figure \ref{fig:manip_b}. The uniform grey region, located at the upper part of the fiber is the place where the film is flat (region A) and gives rise to a zone of regular stripes with a constant wavelength and velocity for the drops (region B). Finally, downstream, some coalescences between drops lead to the formation of a disordered pattern (region C).

\section{Flat film}
\label{sec:flat}

In this section we focus on the region close to the inlet where the film thickness is constant (flat film, grey uniform region, see figure \ref{fig:manip_a}). In the case of Newtonian fluids, the thickness of the film, $h$, is given by the classical Nusselt solution~\cite[]{Duprat2007}. In the case of very thin films ($h \ll R$), i.e. the planar case,  there is a cubic relation between the flow rate on the fiber and $h$. We define the cylindrical coordinates system $(r, \theta, z)$, where $r$ is the radial coordinate (the fiber center is the origin), $\theta$ the azimuthal coordinate and $z$ the axial coordinate oriented downward in the flow direction. In the case of a shear-thinning solution exhibiting normal stress effects, for a steady axisymmetric flow, the stress balance in the axial direction $z$ is written as:

\begin{equation}\label{eq:NS}
	\frac{\partial \sigma_{zz}}{\partial z} + \frac{1}{r} \frac{\partial (r \sigma_{rz})}{\partial r} = \frac{\partial p}{\partial z}-\rho g
\end{equation}

\noindent for $R < r < R+h(z)$, where $\sigma$ denotes the stress tensor, $p$ is the pressure field in the film, $\rho$ and $g$ are respectively the fluid density and the gravitational acceleration.
Since the first normal stress difference, $N_1$, can be expressed as $\sigma_{zz}-\sigma_{rr} = \psi_1 \left( \frac{\partial v}{\partial r}\right)^2$, with $v(r,z)$ the axial velocity, which varies along the film thickness, equation (\ref{eq:NS}) becomes

\begin{equation}\label{eq:NS2}
	\frac{\partial N_1 }{\partial z} + \frac{1}{r} \frac{\partial (r \sigma_{rz})}{\partial r} =\frac{\partial p}{\partial z} -\frac{\partial \sigma_{rr}}{\partial z}-\rho g
\end{equation}

As no spatial variations are detected in the flat film, $z$-invariance of the velocity field implies that $\frac{\partial N_1 }{\partial z}=0$. Thus, normal stress has no effect on the flat film.

To calculate the velocity profile in the flowing film, we model the shear-thinning effect by a power-law in accordance with expression \ref{eq:power_law}.
Thus (\ref{eq:NS2}) becomes

\begin{equation}\label{eq:DE}
	1+\frac{1}{1+\tilde r} \frac{d}{d\tilde r} \left( (1+\tilde r) \left( \frac{d \tilde v(\tilde r)}{d \tilde r}\right)^n \right) = 0
\end{equation}

\noindent where the following dimensionless variables are introduced: $\tilde r = \frac{r-R}{R}$, $\tilde h = \frac{h}{R}$, $\tilde v = \frac{v}{V} $ and $V \equiv R \left( \frac{\rho g R}{\beta} \right)^{1/n}$.

The fluid velocity satisfies two boundary conditions: no-slip on the fiber ($\tilde v(\tilde r = 0) = 0$) and zero tangential stress at the liquid-air interface ($\partial_{\tilde r} \tilde v (\tilde r = \tilde h) =  0$). 
Considering this last boundary condition, equation~(\ref{eq:DE}) becomes:

\begin{equation}\label{eq_to_solve}
	\frac{d \tilde v}{ d \tilde r} = \left( \frac{1}{1+\tilde r} \left( \tilde h(1+\tilde h/2) - \tilde r (1+\tilde r/2) \right) \right)^{\frac{1}{n}}
\end{equation}

The flow rate per unit length $q = \frac{Q}{2 \pi \rho R}$ is written in dimensionless form as 
 \begin{equation}\label{eq:debit}
           \tilde q =q \frac{1}{R V} =  \int_{0}^{\tilde h} \tilde v (\tilde  r +1) d\tilde r 
 \end{equation}

We note that for the Newtonian case $n=1$ and $\beta=\eta$, we recover the analytical Nusselt solution $v=\frac{\rho g}{4\eta} \left( 2(R+h)^2\ln \left(\frac{r}{R}\right) - (r^2-R^2) \right)$.

\begin{figure}
\begin{center}
\includegraphics[scale=1]{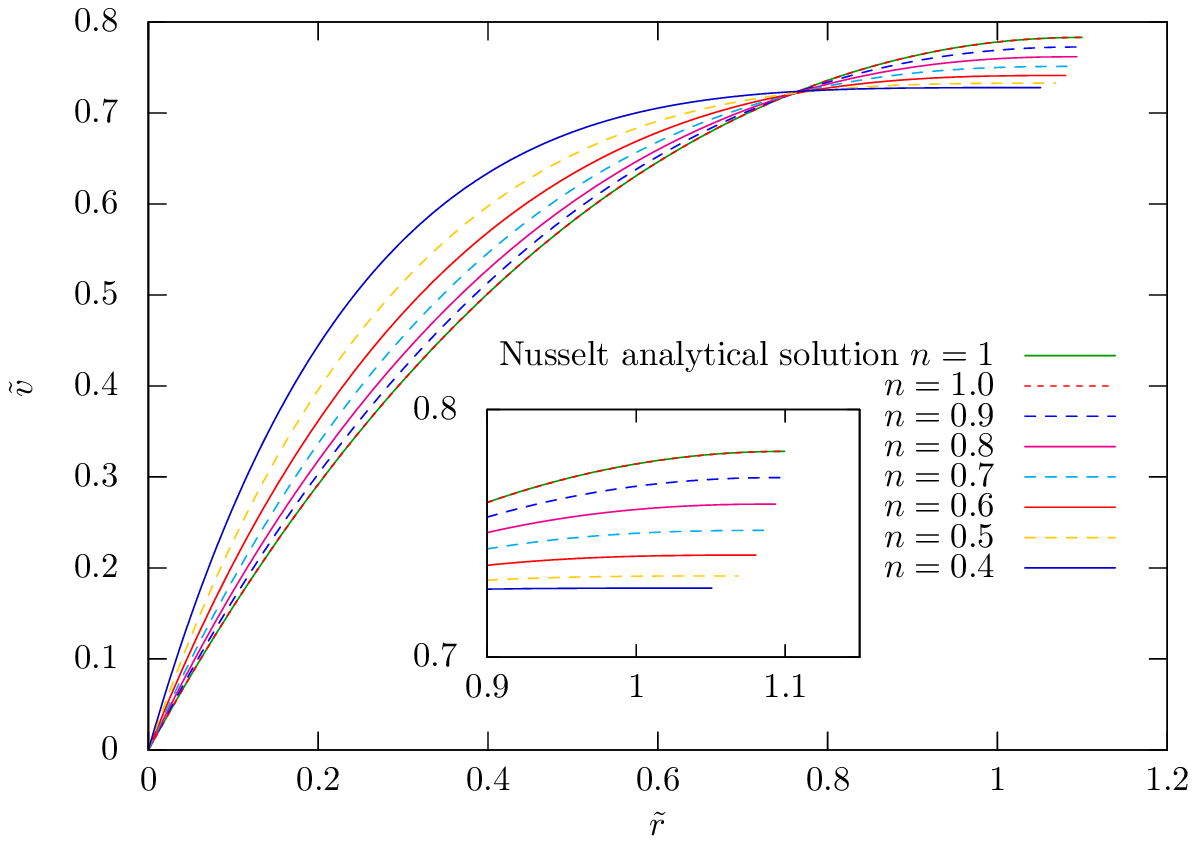}
\caption{Velocity profiles of the flowing film at a constant flow rate ($\tilde q = 1$) for several $n$ values from a Newtonian fluid ($n=1$) to a high shear-thinning effect ($n=0.4$). The inset is a close-up of the liquid-air interface region.}
\label{fig:numerique-v-r}
\end{center}
\end{figure}

\begin{figure}
\begin{center}
\includegraphics[scale=1]{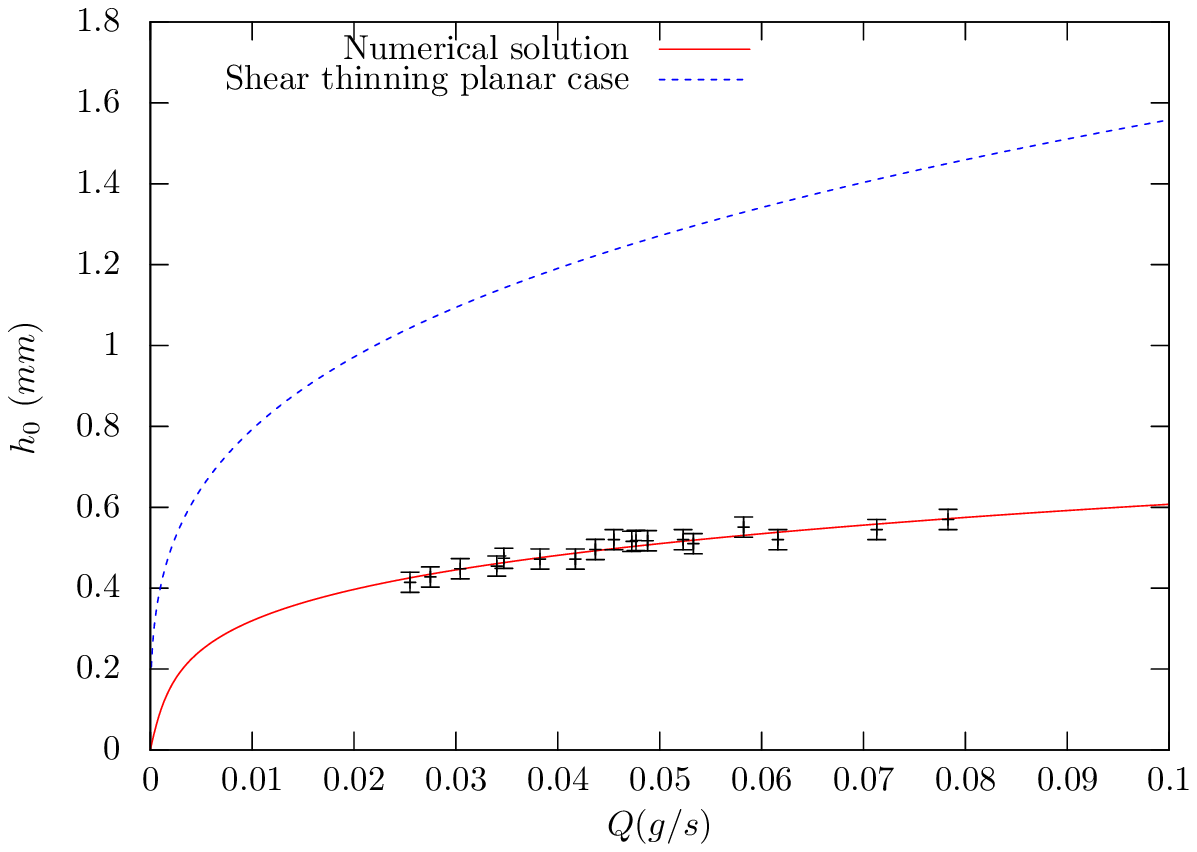}
\caption{ Numerical solution (solid line) and experimental results for a concentration in PAAm equal to $0.4$\%. The dashed curve is the analytical solution for $h\ll R$ given by the equation \ref{eq:debit_plan} for $n=0.73$ .}
\label{fig:experiment-q-h}
\end{center}
\end{figure}

Equation~(\ref{eq_to_solve}) is solved using a Runge-Kutta algorithm (starting at $\tilde v(\tilde r = 0) = 0$ in order to satisfy the boundary condition on the fiber). 
First, the influence of the parameter $n$ is studied for a constant flow rate $\tilde q = 1$, and the integral of the equation~\ref{eq:debit} is estimated by the trapezium rule.
We choose two values for the $\tilde h$ parameter ($h_{\textrm{min}}$ and $h_{\textrm{max}}$) satisfying $\tilde q(\tilde h_{\textrm{min}}) \leq 1 \leq \tilde q(\tilde h_{\textrm{max}})$ and we find the film thickness by a bisection method for which the condition $\tilde q = 1$ is satisfied at $0.1$\%. 
The results are shown in figure \ref{fig:numerique-v-r}. 
The numerical solution for the Newtonian case, $n=1$ is identical to the analytical Nusselt solution.
For a constant flow rate, an increase in the shear-thinning effect modifies the velocity profile shape: the parabolic profile tends to be replaced by a plug-like profile. This results in higher velocity gradient close to the fiber, whereas close to the interface, the velocity gradient is almost zero. The film thickness is always smaller than for Newtonian fluids.
The film thickness is plotted as a function of the flow rate on the film for a PAAm solution (0.4\%,  n=0.71), in figure (\ref{fig:experiment-q-h}). We choose to present experimental data only for PAAm solutions, since for xanthan solutions, the healing length is two or three times smaller than for PAAm solutions. The good agreement between the numerical solutions and our experimental data validates the choice of the Oswald power-law model for the viscosity and also our assumption on the negligible effect of surfactant on the zero-shear stress boundary condition at the liquid-air interface.

\section{Instability growth rate}
\label{sec:insta}

\begin{figure}
\begin{center}
\subfigure[]{\label{fig:lambda_xant}%
\includegraphics[scale=0.8]{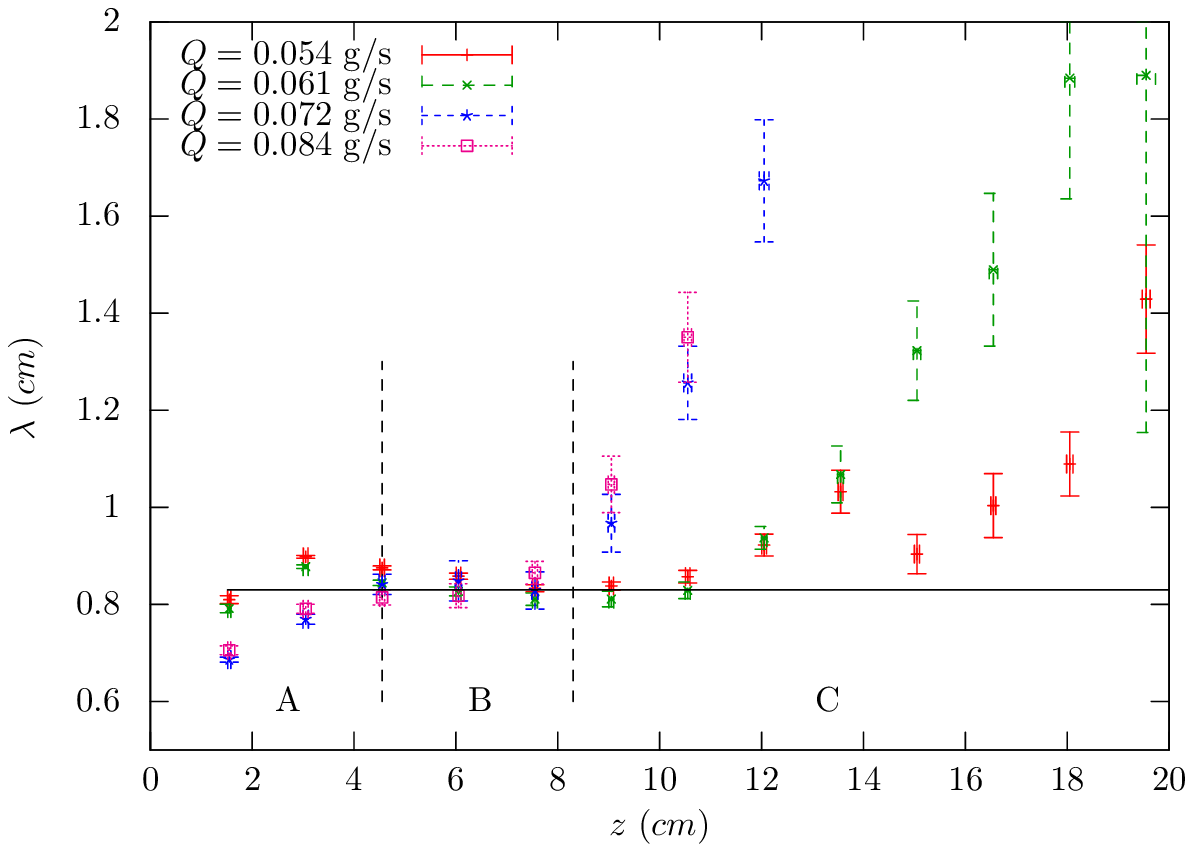}}
\subfigure[]{\label{fig:lambda_paam}%
\includegraphics[scale=0.8]{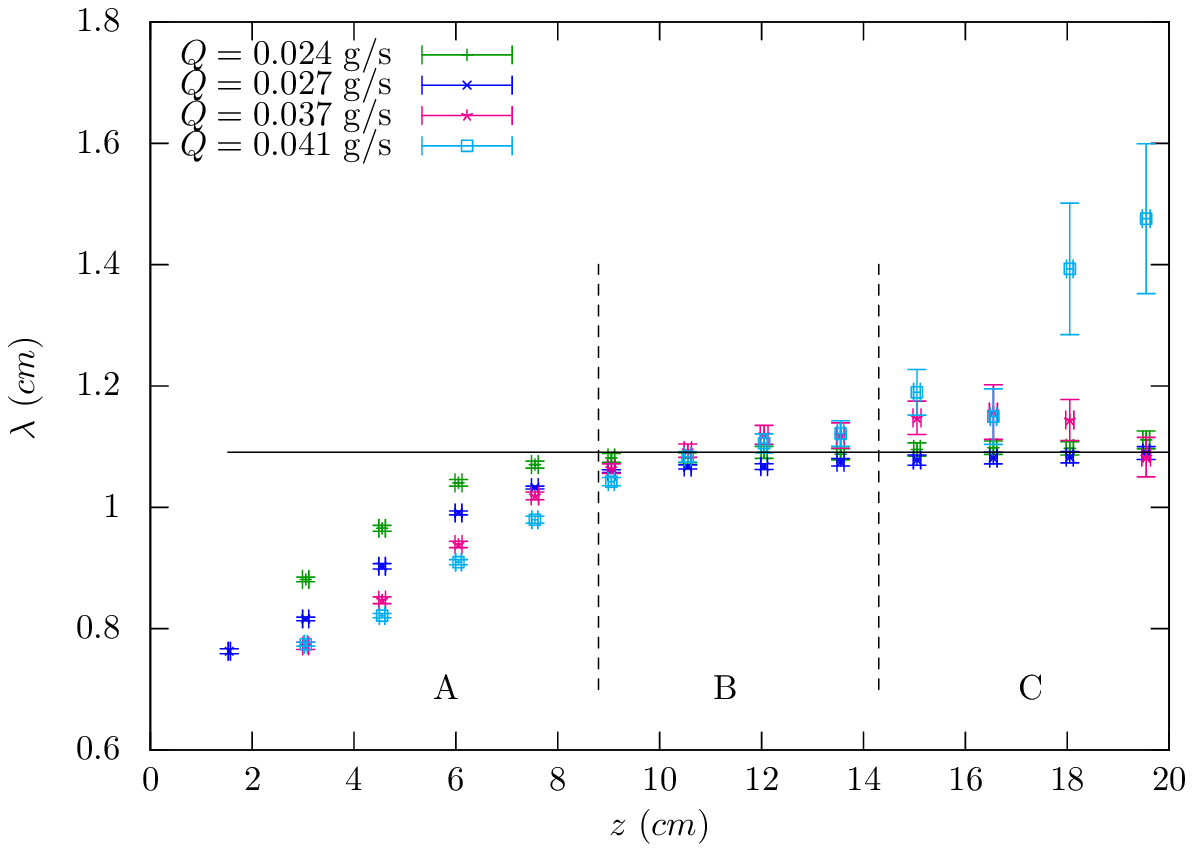}}
\caption{Wavelength at different flow rates for (a) xanthan and (b) PAAm solution. Letters A, B and C denote respectively the growth of the instability, the ordered and the disordered pattern regions.}
\label{fig:lambda}
\end{center}
\end{figure}

The impact of the shear-thinning and elastic effects on the growth rate of the Rayleigh-Plateau instability is investigated through experiments with xanthan (0.8\% NaCl) and PAAm solutions (0.4\%) in order to distinguish the role of each non-Newtonian property.
After the flat film region, some variations on the film thickness are detected and a regular pattern of drops emerges due to the Rayleigh-Plateau instability as shown in figure \ref{fig:manip_c}. 
The wavelength of the drop-like pattern is plotted in figure \ref{fig:lambda} as a function of the distance to the entrance nozzle.
The wavelength increases (regime A in figure \ref{fig:lambda}) until it reaches a well-defined value (regime B in figure \ref{fig:lambda}).
Then, lower down, some coalescence events can disrupt the regular pattern (regime C particularly in figure \ref{fig:lambda}a).
The wavelengths of the regular pattern for non-Newtonian fluids are somewhat higher than those expected with Newtonian fluids. Nevertheless,  in both cases, the classical Rayleigh Plateau wavelength fails to fit the experimental data and is always smaller.
The length of the regular pattern depends on the flow rate (at high flow rates,
coalescence events occur earlier) but  it is typically  is of the order of seven
centimeters for PAAm solutions (figure \ref{fig:lambda}a) and shorter, about
four centimeters, for xanthan solutions (figure \ref{fig:lambda}b). In the former case, we can note that the axisymmetric conformation is not the only case observed on the fiber: some non-axially symmetric conformations can be observed with asymmetric drops. Such conformations have been described by \cite{Carroll1986} as a roll-up transition and must be avoided in our case.

\begin{figure}
\begin{center}
\includegraphics[scale=1]{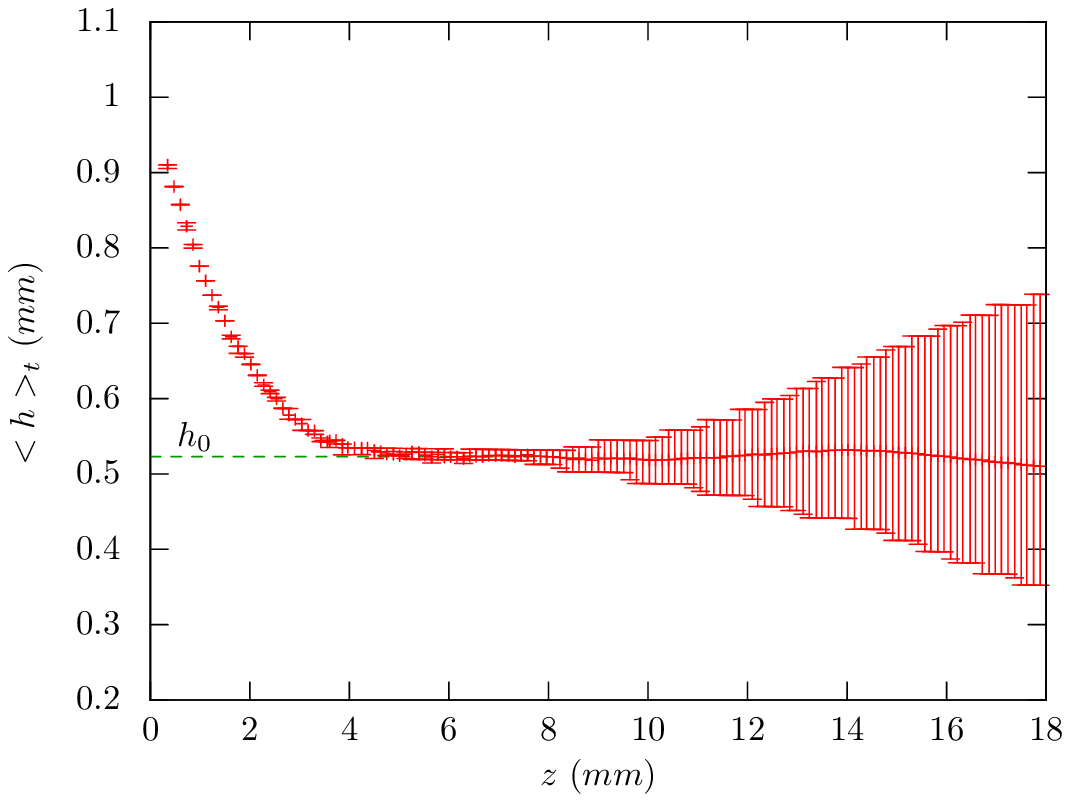}
\caption{Average film thickness in time $<h>_t$ along the $z$ fiber axis ($Q=0.032$ g/s). Bars indicate the extreme values of the film thickness. The flat film thickness is denoted by $h_0$. }
\label{fig:experiment-profil}
\end{center}
\end{figure}

\begin{figure}
\begin{center}
\includegraphics[scale=1]{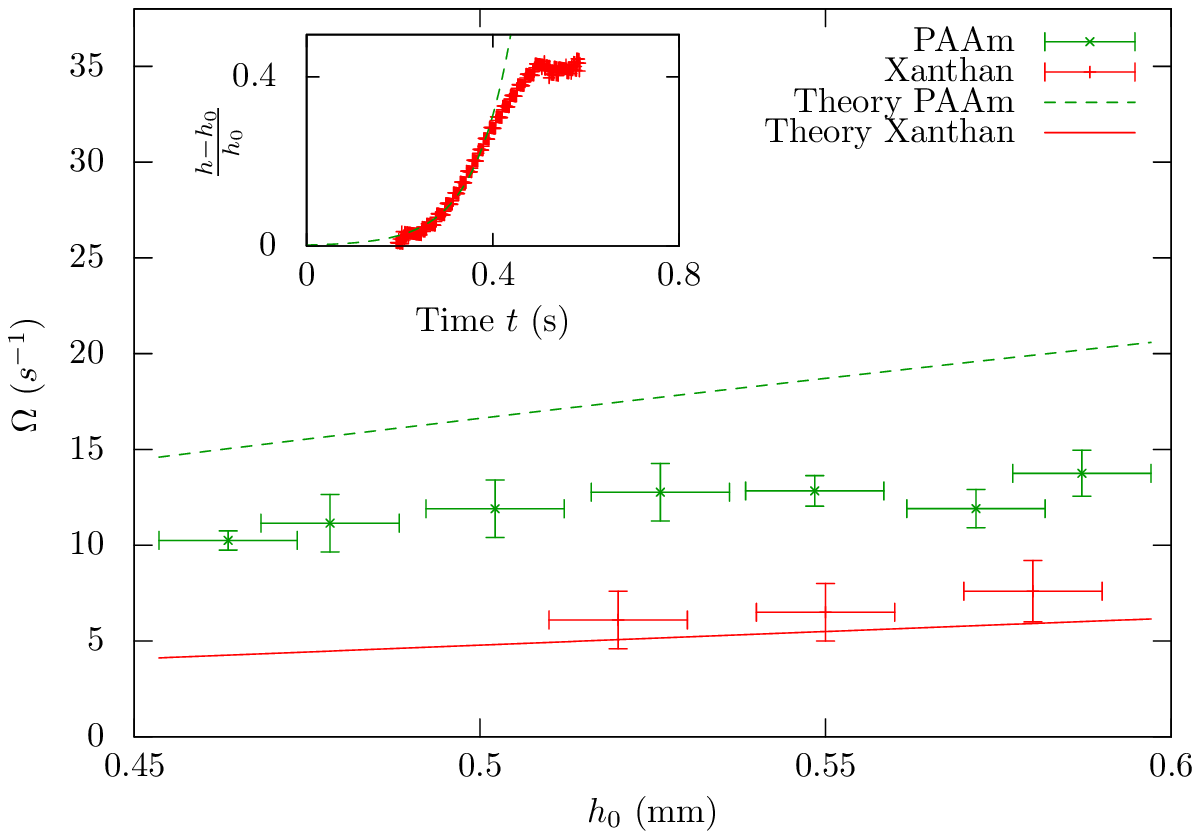}
\caption{The growth rate, $\Omega$, is plotted versus the flat film thickness, $h_0$. The experimental data are represented by the dots and the equation \ref{eq:croissance2} by curves. The inset shows the growth of the film using the method described in section~\ref{sec:insta}.}
\label{fig:croissance}
\end{center}
\end{figure}

In order to characterize the instability growth rate, we record a stack of images at a typical frame rate of 1000 images per second. Then, the position of the film interface is detected over space and time: $h(z,t)$. 
Figure \ref{fig:experiment-profil} shows the average film thickness over the time $<h(z,t)>_t$ and the extremal film positions for a PAAm solution. 
It shows successively the meniscus, the flat film and the onset of the instability which is marked by a strong variation of the film thickness. 
The velocity of the interface is calculated for each stack.
Then, a point on the interface (chosen to become a point of maximum height) is followed at this velocity using the set of data $h(z,t)$. 
The resulting values of the normalized profile $(h-h_0)/h_0$ as a function of time for a typical experiment are plotted in the inset of figure \ref{fig:croissance} and fitted by an exponential law $\frac{h-h_0}{h_0}=Ae^{\Omega t}$ in the early linear stages. 
From this fit, the growth rate, $\Omega$, is extracted and averaged over several other experiments; the results for xanthan and PAAm solutions were reported in figure \ref{fig:croissance}.

A first simplified attempt to obtain an expression for the growth rate consists in a linear stability analysis. The fluid is assumed to exhibit pure shear-thinning effects  with 
 $\eta(\dot\gamma) = \beta \dot\gamma^{n-1}$, and we assume very thin films such that  $h \ll R$ (planar approximation). Thus, in cartesian coordinates, the following momentum equation holds, in the lubrication approximation:

\begin{equation}\label{eq:mvt_plan}
 0 =\Pi + \frac{\partial \eta(\dot\gamma)\dot\gamma}{\partial r} 
\end{equation}

where $\Pi$ is the pressure gradient given by:

\begin{equation}
\Pi =  \rho g +\gamma \left( \frac{\partial_{z} h}{(R+h)^2} + \frac{\partial^3 h}{\partial z^3} \right) \simeq  \rho g + \gamma \left( \frac{\partial_{z} h}{R^2} +  \frac{\partial^3 h}{\partial z^3}\right) \label{eq:pi}
\end{equation}

GIven that there is no fluid slippage on the fiber and no stress on the liquid-air interface, the velocity is calculated from the momentum equation, so that

\begin{equation}
v(r) = \frac{1}{1+1/n}\left( \frac{\Pi}{\beta} \right)^{1/n} \left[  h^{1+1/n} - (h-r)^{1+1/n} \right]
\end{equation}

The flow rate per unit length, defined as $q_p = \int_{0}^{h} v dr$, is given by:

\begin{equation}
q_p = \left( \frac{\Pi}{\beta} \right)^{1/n} \frac{h^{2+1/n}}{2+1/n}
\end{equation}

and also satisfies the mass conservation equation:

\begin{equation}\label{continuite_plan}
\frac{\partial h}{\partial t} + \frac{\partial q_p}{\partial z} = 0
\end{equation}

We assume infinitesimal perturbations around the uniform film thickness, $h_{0}$, and for the corresponding  flow rate, $q_{p0}$, so that

\begin{eqnarray}
h(z,t) = h_0 + h_1(z,t), \qquad  \\
q_p =q_{p0} + q_{p1}, \qquad 
\end{eqnarray}

This results in the following expressions

\begin{eqnarray}
q_{p0} &=& \left( \frac{\rho g}{\beta} \right)^{1/n} \frac{h_0^{2+1/n}}{2+1/n}\label{eq:debit_plan} \\
q_{p1} &=& q_{p0}\left[ \frac{\gamma}{n\rho g}  \left( \frac{\partial_{z} h_1}{R^2} +  \frac{\partial^3 h_1}{\partial z^3}\right)  + \left(2+\frac{1}{n}\right)  \frac{h_1}{h_0}\right]\\
\end{eqnarray}
and in the linearized equation:

\begin{equation}
-\frac{\partial h_1}{\partial t} = \frac{q_{p0} \gamma}{n\rho g} \left( \frac{\partial_z^2 h_1}{R^2} +\frac{\partial^4 h_1}{\partial z^4} \right)+ \left(2+\frac{1}{n}\right) \frac{q_{p0}}{h_0}\frac{\partial h_1}{ \partial z}\label{machin_plan}
\end{equation}

Developing the thickness perturbation as  $h_1 = A e^{i(kz-\omega t) }$ leads to the dispersion relation:

\begin{equation}
\omega(k) =    k \frac{q_{p0}}{h_0} \left( 2 + \frac{1}{n} \right) + i \frac{q_{p0} \gamma}{n \rho g } \left( \frac{k^2}{R^2} - k^4 \right) 
\end{equation}

The maximum of $\Imag (\omega(k))$ gives:

\begin{equation}
\Omega = \frac{q_{p0} \gamma}{4n\rho g R^4}, \qquad h\ll R \label{eq:croissance}\\ 
\end{equation}

As shown in figure \ref{fig:experiment-q-h}, the flow rate given by \ref{eq:debit_plan} does not reproduce the experiment where the planar approximation ($h\ll R$) is not valid. In previous works on Newtonian fluids, it has been shown \cite[]{Craster2009} that for $h\sim R$, the expression for the growth rate is similar except that $R$ should be replaced by $R+h$.
Moreover, we made the choice to use our numerical calculation described in section \ref{sec:flat} which provides $q_{\textrm{num}}(h)$ without assumption on the film thickness. 
Finally, the addition of a large amount of surfactant modifies the growth rate by a factor $4$, as described by \cite{Carroll1974} since surfactants change the surface elasticity.

Taking into account these corrections, we obtain the expression
\begin{equation}
\Omega = \frac{1}{4}\frac{q_{\textrm{num}}(h) \gamma}{4n\rho g (R+h)^4}\label{eq:croissance2}\\ 
\end{equation}

which is plotted in figure \ref{fig:croissance} for both chemical systems.

\begin{figure}
\begin{center}
\includegraphics[scale=0.7]{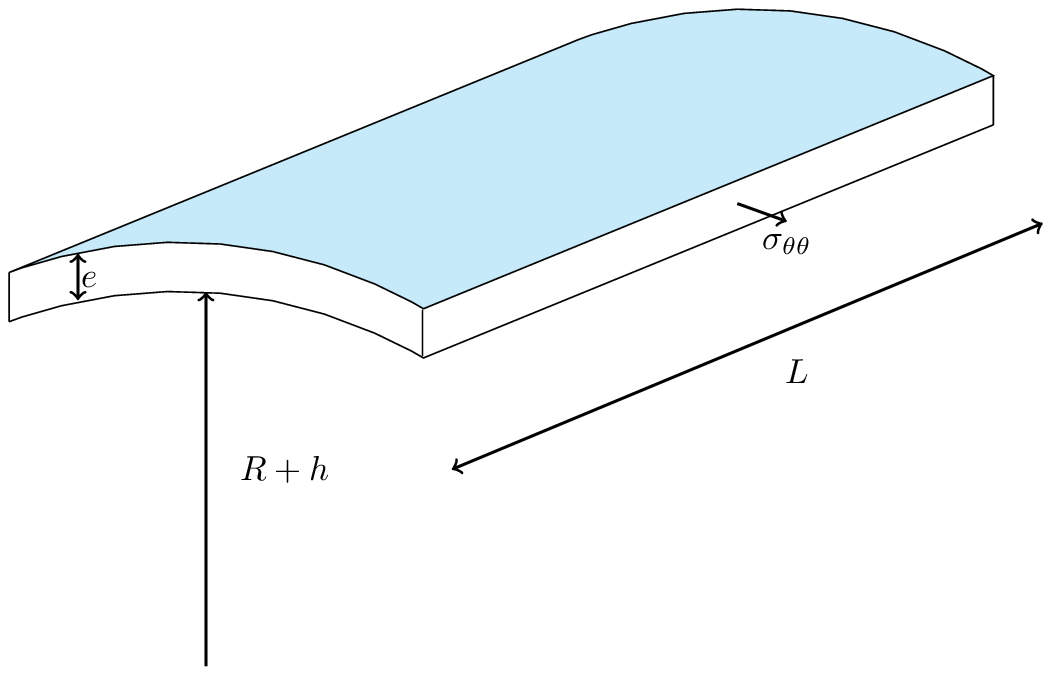}
\caption{Stretching stress, $\sigma_{\theta\theta}$, in a thin cylindrical shell of radius $R+h$ and thickness $e\ll R+h$.}
\label{fig:hoop_stress}
\end{center}
\end{figure}

Data for the xanthan solution are well described by \ref{eq:croissance2}. To validate our method and to compare with a Newtonian fluid of similar surface tension ($\gamma=20.9$ mN/m) and viscosity ($\eta = 0.965$ Pa.s), we have performed an experiment with a silicon oil (with $\rho=96.5$ kg.m$^{-3}$ and $h_{0}=0.55$ mm) and measured a growth rate $\Omega$ equal to $10.9 \pm 0.7$ s$^{-1}$. Equation \ref{eq:croissance2} for $n=1$ gives $\Omega$ equal to $11.1$ s$^{-1}$. The small variations in growth rate between the xanthan solution and the silicon oil are reasonable as the value of the viscosity is of the same order. 
Concerning  PAAm solutions,  there is significant deviation from the theory due to the normal stress of this solution. A qualitative explanation of the role of normal stress can be provided by the ``hoop stress'' effect \cite[]{Graham2003}.
A short description of this effect can be made by considering the liquid surface as an infinite cylindrical shell of thickness $e$  (figure \ref{fig:hoop_stress}). For a cylinder of radius $R+h$, the balance between the internal pressure  $P$ and the stretching stress $\sigma_{\theta\theta}$, leads to $ 2(R+h) L P= 2 L \sigma_{\theta\theta} e$. The internal pressure $P$ is generated by the normal stress $N_1$ in the bulk.
Interpreting the stretching force per unit length, $2\sigma_{\theta\theta} e$, in term of surface tension $\gamma_{\psi_1}$, we obtain

\begin{equation}
 \gamma_{\psi_1} = -\psi_1 \dot\gamma^2 (R+h)
\end{equation}

An estimation for $\dot\gamma=100$ s$^{-1}$ gives a $\gamma_{\psi_1}$ of about $-10$ mN/m significantly lowering the effective surface tension. The growth rate should therefore be estimated with the effective surface tension lower than the fluid surface tension and the resulting curve for the growth rate would be shifted and enable us to recover the experimental data. 

\section{Drop morphologies: normal stress effect}
\label{sec:drop}

\begin{figure}
\begin{center}
\includegraphics[scale=1]{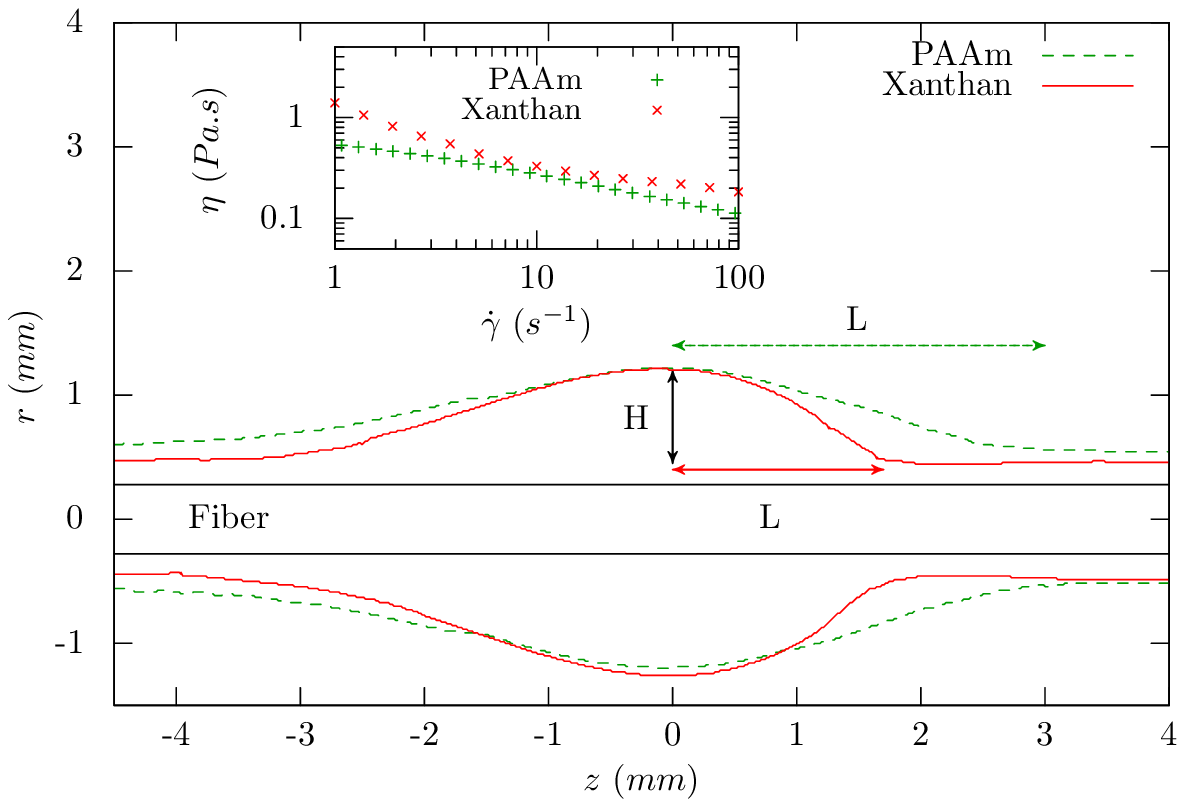}
\caption{Drop shapes for two polymer solutions. The drop front of the viscoelastic solution (dashed green line) is swollen compared with the pure shear-thinning liquid (solid red line).}
\label{fig:profil_goutte}
\end{center}
\end{figure}

This section is devoted to a comparison between the patterns of flowing films of PAAm and xanthan, focusing particularly on the normal stress effect on the shape of the drops.
Such a comparison requires polymeric solutions having similar shear-thinning properties.
Further, $0.8$\% NaCl was added to xanthan ($0.4$\%) solutions to decrease the high shear-thinning effect (figure \ref{fig:xanthan}). 
Optimal adjustment of the shear-thinning of PAAm solutions ($0.6$\%) is achieved as shown in the inset of figure \ref{fig:profil_goutte}. Thus, the difference between the two solutions concerns only the presence or absence of normal stress.

The typical pattern observed on the fiber consists of an axisymmetric film of constant thickness.
Then the film breaks up spontaneously into a drop-like wave train as described in the previous section.

\begin{figure}
\begin{center}
\includegraphics[scale=0.6]{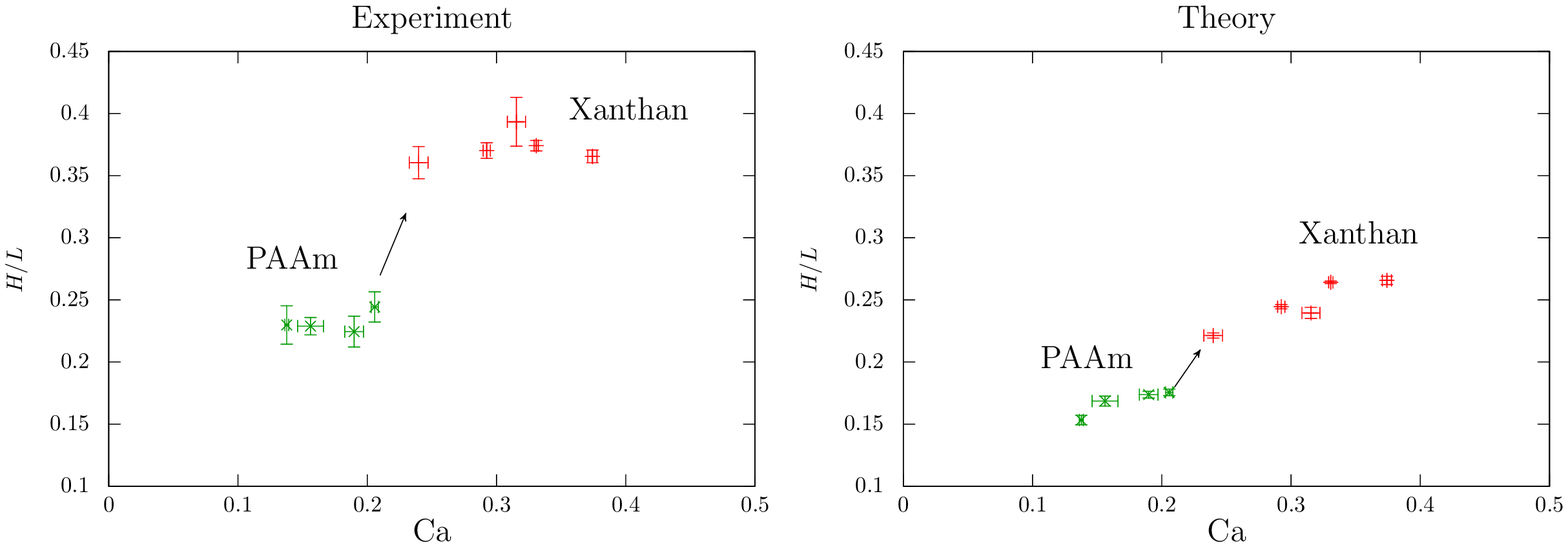}
\caption{Slopes of drop fronts for xanthan and PAAm as a function of the capillary number $\textrm{Ca}$.}
\label{fig:pente_goutte}
\end{center}
\end{figure}

For axisymmetric patterns, the superposition of drop profiles in a PAAm and xanthan films is shown in figure \ref{fig:profil_goutte}.
Clear differences can be noticed in the profiles, notably the steepening of the drop front for the PAAm solution compared to the xanthan solution. In both profiles, there is a clear asymmetry between the front and back of the drops which is more accentuated for the xanthan drop. This remark suggests that the shape of the drops is affected by gravity.
The apex heights of both drops are identical as well as the film substrate between drops (the trailing edge for PAAm is longer than for xanthan).
These observations confirm the fact that the normal stress plays a significant role in the thin regions, close to the tail and the front of the drops, but exhibits no effect in the center of the drop (the thick region).
To quantify experimentally the swelling effect observed with PAAm solution, we define the slope of the front $H/L$ as shown in figure \ref{fig:profil_goutte}.
This parameter is plotted in PAAm and xanthan solutions for different flowing rates in figure \ref{fig:pente_goutte}.

To highlight the swelling process, we consider a scaling law analysis, starting from the stress balance \ref{eq:NS2}.
Since the film is not flat the $z$-invariance is no longer valid.
Exhibiting the contribution of the normal stress difference $\sigma_{zz}-\sigma_{rr}$, and shear stress $\sigma_{rz}$, equation \ref{eq:NS2} becomes

\begin{equation}\label{eq:NS3}
	\frac{\partial (\psi_1 \left( \frac{\partial v}{\partial r}\right)^2)}{\partial z} + \frac{1}{r} \frac{\partial (r \eta(\dot\gamma) \dot\gamma)}{\partial r} =\frac{\partial p}{\partial z} -\frac{\partial \sigma_{rr}}{\partial z}-\rho g
\end{equation}

The axial velocity $v(r,z)$, is determined using a series of the form \cite[]{Boudaoud2007}:

\begin{equation}
 v(r,z) = a_0(z) + a_1(z) r + a_2(z) r^2
\end{equation}

Functions $a_0(z)$, $a_1(z)$ et $a_3(z)$ are calculated using the following three equations, two for the boundary conditions and the last one for the mass conservation:
\begin{itemize}
 \item Boundary condition at the interface with the fiber $ v(r=R,z)=0$;
\item Boundary condition at the liquid/air interface $\frac{\partial v(r=R+h(z))}{\partial r}=0$;
\item The mass conservation equation,
\end{itemize}

\begin{equation}
 \frac{\partial h}{\partial t} + \frac{\partial}{\partial z} \int_R^{R+h(z,t)} v(r,z) dr=0
 \label{eq:flowrate}
\end{equation}

Considering equation \ref{eq:flowrate} in the reference frame of a drop moving at a velocity $U$ and using the condition that the mean flow rate satisfies $\overline{q}\sim Uh \underset{h\rightarrow 0}{\longrightarrow}0$, the axial velocity is 

\begin{equation}
  v(r,z) =  \frac{3}{2} U \frac{(R-r)(r-2h(z)-R)}{2h^2(z)}
\end{equation}

Assuming a constant viscosity, $\eta$, the stress balance equation \ref{eq:NS3} to zero-order in $r$ is

\begin{equation}\label{eq:NS4}
	3 \eta U \frac{h(z)-R}{2R h^2(z)} - U^2 \psi_1 \frac{9 h' }{2h^3(z)}  =\frac{\partial p}{\partial z} -\frac{\partial \sigma_{rr}}{\partial z}-\rho g
\end{equation}

The normal stress balance at the free surface of the film assumes that  $-P+\sigma_{rr}=\gamma \kappa$ with $\kappa$ the curvature of the interface. 
If $L$ is the characteristic length in the axial direction and if $H$ is the characteristic apex height of the drop, then the normal stress balance at the free surface of the film, accounting for the curvature, is given by

\begin{equation}\label{eq:NS5}
\frac{\partial p}{\partial z} -\frac{\partial \sigma_{rr}}{\partial z}  \sim -\gamma \left( \frac{\partial_{z}h}{R^2} + \frac{\partial^3 h}{\partial z^3}\right)
\end{equation}

Thus the right-hand side of equation \ref{eq:NS4} becomes:

\begin{equation}\label{eq:NS6}
\frac{\partial p}{\partial z} -\frac{\partial \sigma_{rr}}{\partial z}-\rho g \sim -\gamma \left( \frac{H}{L^3} + \frac{H}{LR^{2}}+l_{c}^{-2}\right)
\end{equation}

where $l_{c} = \sqrt{ \frac{\gamma}{\rho g}}$ is the capillary length.

Experimental observations suggest that $\frac{H}{L^3} \ll \frac{H}{LR^2}$ and $ \kappa^{2} \ll \frac{H}{LR^2} $.
So, the scaling law analysis leads to the following equation:

\begin{equation}
\frac{H^3}{L^3} \sim  \frac{R^{2}}{ L^2}\textrm{Ca}  \left(1-6  \frac{\mathscr L}{L} \right)
\label{eq:formegoutte_echelle}
\end{equation}

where $\textrm{Ca}=\frac{\eta U }{\gamma}
$ is the capillary number and $\mathscr L = \frac{\psi_1 U}{\eta}$ is  the normal stress characteristic length.

This scaling law gives the slope of the front of a drop, $H/L$, as a  function of the viscoelastic properties of the polymeric solution.
In particular, in the case of polymeric solutions exhibiting normal  
stress, $\psi_1 \neq 0$, the expression \ref{eq:formegoutte_echelle}  
clearly shows that $H/L$ decreases.
A comparison between the experimental data and the results of the scaling analysis is presented in figure   
\ref{fig:pente_goutte} for different flow rates. There is a good agreement between the experiment and the model which succeeds in highlighting the swelling effect on the drop shape induced by the normal stress effect.

\section{Conclusion}

The effects of non-Newtonian properties of fluids have been investigated in  
the case of a film flowing down a vertical fiber.
The flow on the fiber can be divided into three regions: (A) at the inlet, the film exhibits a uniform  
thickness, i.e the flat film region; (B) the uniform film is progressively replaced by a well-defined pattern of drops separated by a thin film, i.e the Rayleigh-Plateau region; (C) 
the coalescence of drops disrupts the flow and give rise to a disordered pattern.
In order to disentangle the role of the shear-thinning effect and of the normal stress, we have considered two kinds of polymer solutions.
The first consists of rigid rod-like polymers (xanthan), exhibiting  a strong shear-thinning behaviour but negligible elastic effects. 
For the second solution, we used flexible polymers (PAAm) exhibiting strong elastic effects and shear-thinning effects similar to those of xanthan under certain physico-chemical conditions.
Some adjustments have been made by modifying both the polymer concentration and the physico-chemical properties of the solutions to enhance or reduce one of the non-Newtonian properties : shear-thinning or elastic effect.

Consequences of both effects have been investigated in the first two regions of the flowing film.
In the flat film region, due to the invariance of the film thickness in the axial direction, only the shear-thinning effect is effective. 
At a constant flow rate, our experiments demonstrate that, as a consequence of the shear-thinning effect,  the thickness of the film is always smaller than in the case of Newtonian fluid. 
Our results clearly show the influence of the shear-thinning effect on the velocity profile: a parabolic profile in the Newtonian case tends to 
become a plug-like profile. Thus, an increase of the shear-thinning effect yields a thinner, unperturbed film. 
Further downstream on the fiber, the film undergoes the Rayleigh-Plateau instability.
The growth rate of the instability has been investigated experimentally and theoretically 
using a linear stability analysis. Good agreement is found between the experimental data for xanthan and the model. For PAAm solutions, and to take into account the normal forces, we consider an effective surface tension (lower than the fluid surface tension) which tends to decrease the growth rate and to recover the experimental data.
The morphology of the patterns resulting from the instability depends on the non-Newtonian properties. In particular, the drops formed with PAAm solutions exhibit a swelling effect compared to drops observed with xanthan solution, for a similar shear-thinning effect. 
We observe that the drop of fluid with normal forces is less rounded compared with the case of a pure shear-thinning drop. This swelling effect has been quantified by a scaling law analysis where the slope of the drop front is expressed as a function of the normal stress.

As a conclusion, by considering two kinds of polymeric solutions with the same shear-thinning effect, which differ from each other in the presence of normal forces, we have succeeded in understanding the relationship between  
the rheological properties and the destabilization of the flowing film on a fiber as well as the morphology of the observed patterns.
This should be helpful in understanding what happens with more complex fluids, in particular fluids which exhibit more elastic effects where the elasticity could prevent the growth of the instability.

\paragraph{Acknowledgments:}
We thank F\'ed\'eration Paris VI (high-speed camera) and Triangle de la Physique  
(rheometer apparatus). The authors thank Liyan Yu and Prof. John Hinch for  
fruitful discussions. Also we thank Lionel Auffray, Rafael Pidoux and Alban Aubertin 
for the experiment engineering and technical improvements.


\bibliography{biblio}
\end{document}